\documentstyle[12pt]{article}
\newcommand{\singlespace}{
    \renewcommand{\baselinestretch}{1}\large\normalsize}
\newcommand{\doublespace}{
    \renewcommand{\baselinestretch}{1.6}\large\normalsize}

\newcommand{\be}{\begin{equation}}
\newcommand{\ee}{\end{equation}}
\newcommand{\ba}{\begin{eqnarray}}
\newcommand{\ea}{\end{eqnarray}}
\newcommand{\ket}[1]{| {#1} \rangle}
\newcommand{\bra}[1]{\langle {#1} |}
\newcommand{\ave}[1]{\langle {#1} \rangle}
\newcommand{\bsigma}{\mbox{\boldmath $\sigma$}}
\newcommand{\btau}{\mbox{\boldmath $\tau$}}

\begin{document}
\begin{titlepage}
\pagestyle{empty}
\vspace{1.0in}
\begin{flushright}
SUNY-NTG-94-54
\end{flushright}
\begin{flushright}
October 1994
\end{flushright}
\vspace{1.0in}
\begin{center}
\doublespace
\begin{large}
{\bf{WHERE THE NUCLEAR PIONS ARE}}\\
\end{large}
\vskip 1.0in
G.E. Brown, M. Buballa, Zi Bang Li\\
{\small
{\it Department of Physics, State University of New York,\\
Stony Brook, New York 11794, U.S.A.}}\\
and\\
J. Wambach\\
{\small
{\it Physics Department, University of Illinois,\\
Urbana, IL61801, U.S.A.}\\
and\\
{\it IKP (Theorie), Forschungszentrum J\"ulich\\
D-52425 J\"ulich, Fed. Rep. Germany}}
\end{center}
\vspace{2cm}

\begin{abstract}
Three recent experiments, which looked at pionic effects in nuclei
have concluded that there are {\it no excess pions}. This puts into serious
question the conventional meson-exchange picture of the nucleon-nucleon
interaction. Based on arguments of partial restoration of chiral symmetry
with density we propose a resolution to this problem.
\end{abstract}
\end{titlepage}

\doublespace

\section{Introduction}

\setcounter{equation}{0}

In an article entitled ``Where are the Nuclear Pions?'',
Bertsch {\it et al} \cite{bfsg} discussed three recent experiments
which looked for pionic effects in nuclei. One is a $(\vec p,\vec n)$
quasielastic polarization transfer experiment at LAMPF which,
more or less, directly determines the ratio of spin-longitudinal to
spin-transverse response functions in the nucleus.
At energies below the quasielastic peak conventional models predict
this ratio to be significantly larger than
unity, while experiment finds a ratio slightly below one. This puts
into focus the strength of the tensor interaction in nuclei.
In fact, experiment would suggest that at
the measured momentum transfer $V_{\rm tensor}\sim 0$ while for free
nucleon-nucleon  interactions it is  large.
The second is a new deep inelastic
muon scattering experiment \cite{apma} which no longer sees a significant
enhancement in the EMC ratio $F_2^{A}/F_2^{D}$ for $0.1 \leq x\leq 0.3$.
Since this region is most sensitive to the virtual
pion field in the nucleus it was concluded that there are {\it no excess
pions}.
Very recently it was realized that the enhancement of the pion field
was strongly overestimated in the past
because of incorrectly normalized wave functions.
Introducing the proper normalization factors there is no contradiction
between a conventional model calculation and the measured EMC ratio anymore
in the kinematical regime which is dominated by the pion \cite{MB}.
However, the normalization factors remove only about half of the discrepancy
found between a recent Drell-Yan experiment at Fermi Lab \cite{adml}
and the theoretical predictions.
With appropriate choice of kinematics, this experiment
directly probes modifications of the sea quarks in the nucleus and is therefore
more sensitive to the pion field than the deep inelastic scattering experiment.
A pion excess in nuclei would predict a strong $A$-dependence of the Drell-Yan
ratio and none was observed.
Together with the $(\vec p, \vec n)$ data this calls into
serious question the conventional meson-exchange picture of the nuclear
interaction.

The authors of ref.~\cite{bfsg} suggest that the answer might be
found in the modification of gluon properties in the nucleus, suppressing
the pion field at distances below 0.5 fm. In the present communication
we shall argue that a reasonable explanation lies elsewhere, namely in the
partial restoration of chiral invariance with density. Basically, our
explanation will involve the fact that at finite density the hadronic
world is ``swelled''. More precisely, masses of hadrons made up out of
light up and down quarks decrease with density, all at about the same
rate \cite{brgm} thus:
\be
\frac{m_{N}^{*}}{m_{N}} \cong \frac{m_{\rho}^{*}}{m_{\rho}}
\cong \frac{m_{\omega}^{*}}{m_{\omega}} \cdots \cong \frac{f_{\pi}^{*}}
{f_{\pi}} ,
\ee
where $f_{\pi}$ is the pion decay constant, but here its
more relevant meaning is that of the order parameter for chiral
symmetry breaking. There are approximate signs in the equalities
because these were shown \cite{brgm} to hold only at mean field level,
and loop corrections will enter in higher order. The mass of the pion
$m_\pi$ is exempted from this scaling, because it originates from a
higher scale than QCD, possibly the electroweak scale. In fact, from the
study of pionic atoms \cite{ewtw} we know that due to
many-body effects, the pion mass increases slightly, by
$\sim 5$ MeV, in going to the saturation density, $\rho_0$, of nuclear
matter.

To proceed, we should recall some recent developments in the description
of the nucleon-nucleon potential. Some time ago Thomas \cite{tawh} showed
that the
data on the sea quark content of the proton can be used to obtain
restrictions on the $t$-dependence of the $\pi NN$ vertex function
$\Gamma_{\pi NN}$. Frankfurt, Mankiewicz and Strikman \cite{fmsl} extended this
analysis, finding that the cut-off, $\Lambda_\pi$, in a monopole
parameterization of $\Gamma_{\pi NN}$ should be less than 0.5 GeV. Inclusion
of more mesons in the nucleon cloud allowed Hwang, Speth and Brown \cite{hsbw}
to raise this value to $\sim$ 950 MeV. Clearly such values are too low to
 correctly describe the NN-scattering data and the binding properties of
the deuteron. In various versions of the Bonn potential, $\Lambda_\pi$
is typically 1.2-1.3 GeV. In an effort to reconcile the deep-inelastic
scattering data on the proton with the two-nucleon properties Holinde and
Thomas \cite{htka} chose $\Lambda_\pi=0.8$ GeV but had to introduce an
additional
pseudoscalar meson (which they call $\pi'$) of mass 1.2 GeV. Assuming a
hard form factor of $\Lambda_{\pi'}=2$ GeV, the $\pi'NN$ coupling constant
was adjusted to fit the $NN$ data. More recently it has been recognized
\cite{jdhp} that there are at least two objects with pionic quantum
numbers, one
the elementary pion and the other the correlated $(\rho\pi)$-system
coupled to the quantum numbers of the pion. The latter may explain the
properties of the Holinde-Thomas $\pi'$ \cite{htka}. We find this scenario
quite convincing and shall employ the  Holinde-Thomas ({\it HT}) interaction
in our calculations.

\section{Theoretical Development}

\setcounter{equation}{0}

The enhancement of the pion field is driven by the longitudinal spin-isospin
part of the $NN$ interaction \cite{edam}.
Employing the {\it HT} interaction \cite{htka} it is given by
\be
V_\parallel({\bf q},\omega)=[V_\pi(q,\omega)+V_{\pi'}(q,\omega)]
\bsigma_1\!\cdot\!\hat{\bf q}\bsigma_2\!\cdot\!\hat{\bf q}\,
\btau_1\!\cdot\!\btau_2.
\ee
where
\be
V_\pi(q,\omega)={f_{\pi NN}^2 \over m_\pi^2}{\Gamma_\pi^2(q,\omega)
q^2 \over \omega^2-\left(q^2+m_\pi^2\right)}
\ee
and similarly for the $\pi'$ meson. In the nuclear medium, this interaction
acquires an additional repulsive contribution, usually expressed by
the Migdal parameter $g'$:
\be
\tilde V_\parallel=V_\parallel+{f_{\pi NN}^2\over m_\pi^2}g'_{NN}
\bsigma_1\!\cdot\!\hat{\bf q}\bsigma_2\!\cdot\!\hat{\bf q}\,
\btau_1\!\cdot\!\btau_2
\ee
due to short-range correlations induced by the core for the $NN$ potential.
As Baym and Brown have shown \cite{bbgg} $g'_{NN}$ receives a significant
contribution
from $\rho$-meson exchange, which generates the spin-isospin transverse
interaction,
$V_\perp$. This observation will be important to our discussion. The Migdal
parameter $g'$ can be calculated by a momentum-space convolution of the
central part of the spin-isospin interaction $(V_{\rm central}=
1/3(V_\parallel+2V_\perp)$
with a two-nucleon correlation function $g$:
\be
\tilde V_{\rm central}({\bf k},\omega)=\int {d^3k\over (2\pi)^3}g({\bf k}
-{\bf q})V_{\rm central}({\bf q},\omega) .
\ee
To a good approximation $g({\bf q})=(2\pi)^3\delta({\bf q})-(2\pi^2)
\delta(|{\bf q}|-q_c)/q^2$, where $q_c$ is of the order of the omega-meson
mass ($q_c=3.94$ fm$^{-1}$ \cite{abmg}). The resulting $g'_{NN}$ is $\omega$
and
$q$ dependent and is displayed in the static limit ($\omega=0$) as
the full line in Fig.~1. For small $q$ the value is somewhat
lower than those extracted from Gamow-Teller systematics \cite{ostf}
$g'_{NN}=0.7-0.8$).
On the other hand it agrees well with G-matrix calculations \cite{bbns}.
The increase in $g'_{NN}$ at larger $q$ arises from the $\rho$-meson
exchange tensor interaction.

Another important ingredient in the description of the virtual pion
field is the strong pionic p-wave coupling of the nucleon to $\Delta$(1232)
isobar. The corresponding spin-isospin correlations are described via
transition potentials of the form (2.1) with suitable modifications for the
coupling constants and spin-isospin operators. Also in this case,
short-range  correlations have to be included.
Thies \cite{tmhi} pointed out that $g'_{N\Delta}(0)$ must be close to the
classical Lorentz-Lorenz value 1/3 in order to explain the absence of
multiple scattering in pion-nucleus interaction. Johnson \cite{jmoh}
finds $g'_{\Delta\Delta}(0) = 0.40 \pm 0.13$ which is also consistent with
the classical Lorentz-Lorenz value. Therefore we choose $q_c=8.66 fm^{-1}$ so
that for the $NN\to N\Delta$ and $N\Delta\to N\Delta$ transition potentials
we reproduce  $g'_{N\Delta}(0) = g'_{\Delta\Delta}(0) = 1/3$.
The resulting momentum dependence is also
indicated in Fig.~1. The physical origin of the difference in $g'_{NN}$,
$g'_{N\Delta}$ and $g'_{\Delta\Delta}$ lies in the role of the Pauli
principle as was shown by Delorme and Ericson \cite{jdme} and by
Arima {\it et al} \cite{acsh} some time ago.
Given $\tilde V^{NN}_\parallel$ and the corresponding transition potentials
all pionic properties, relevant to
the experiments under discussion, can be evaluated in linear response
theory within the Random-Phase-Approximation. This is the standard
scenario employed by many people.

With dropping masses there are several modifications. In medium, the
nucleon acquires an effective mass,  the mass that enters into the
quasiparticle velocity
\be
v_{QP}= {p\over m_N^*},
\ee
where $p$ is the quasiparticle momentum. By itself this is not unconventional
and it is often incorporated in the standard treatment. The crucial point,
as was shown by Brown and Rho \cite{brgm}, is that $m_N^*$ is related to the
chiral order parameter $f_\pi^*$ as
\be
{m_N^* \over m_N} \cong \sqrt{g_A^*\over g_A}{f^*_\pi\over f_\pi}
\ee
once loop corrections are included which bring in the axial vector
coupling constant ($g^*_A$ denotes the in-medium
coupling constant). The scaling relation
$m^*_\rho/m_\rho=f_\pi^*/f_\pi$
then directly links the in-medium mass of the $\rho$ meson to $m^*_N$ as
\be
{m^*_\rho \over m_\rho} =\sqrt{g_A\over g_A^*}
{m_N^* \over m_N}.
\ee
A drop of the rho-meson mass with density will increase the range of
the spin-transverse interaction, $\tilde V_\perp$, as well as its coupling
constant.  We note that \cite{bbns}
\be
{f_{\rho NN} \over m_\rho} =g_{\rho NN} {\left( 1+\kappa_V^\rho \right)
\over 2m_N},
\ee
where $\kappa_V^\rho$ is the anomalous $\rho$-meson tensor coupling to the
nucleon. In accordance with \cite{brgm} $g_{\rho NN}$ will not depend
on density which follows naturally if one considers the $\rho$
meson as the gauge particle of the hidden symmetry \cite{bkym}.
With (2.6) we then obtain that
\be
{f_{\rho NN}^* \over f_{\rho NN}}=\sqrt{g_A\over g_A^*},
\ee
the $f^*_{\rho NN}$ being the in-medium coupling. There will also be a
change in the $\pi NN$ coupling constant $g_{\pi NN}$.
In analogy to (2.8) we have
\be
{f_{\pi NN}\over m_\pi}={g_{\pi NN}\over 2m_N}.
\ee
As noted, $m_\pi$ changes but little with density, increasing by
$\sim 5$ MeV for $\rho\sim \rho_0$ \cite{ewtw}. In chiral perturbation
theory, a change in $f_{\pi NN}$ enters first in one-loop calculations,
four powers higher in the Weinberg counting rules \cite{wsei} than the
basic pion exchange interaction. Consequently, changes in $f_{\pi NN}$ with
density are expected to be small and we neglect them. Thus, the ratio
$f_{\pi NN}/m_\pi$ is taken not to change with density. From the
Goldberger-Treiman relation which can be written as
\be
{g_{\pi NN}\over m_N} = {g_A\over f_\pi} = {2f_{\pi NN}\over m_\pi}
\ee
this requires $g^*_{\pi NN}$ to scale as $m^*_N$ with density. This
will turn out to be quite important for the deep inelastic experiments.

The cut-off parameter $\Lambda_\pi$,
which determines the extent of the pion in $\pi$-nucleon interactions and
boson exchange models will also be modified in the medium.
We can see this in the following way.
Consider the $\pi NN$ vertex and the mass dispersion in the pion channel.
Structures, other than elementary pion, will set in with the correlated
($\rho,\pi$)-state (see Fig.~2).
The mass $\Lambda_\pi$ involved in this new structure will be determined
by the integral which involves cutting horizontally through the
$\rho$ and $\pi$ lines, and putting them on shell:
\be
\Lambda_\pi \sim m_\rho + m_\pi + T_{\rho \pi}
\ee
where $T_{\rho \pi}$ is the summed kinetic energy of the $\rho$
and $\pi$. Carrying out the integral in the dispersion relation
self-consistently will involve inserting the $\pi NN$ vertex function
appropriately. This has been done many times,
most recently by Janssen {\it et al} \cite{jdhp}. The net result is that
\be
{\Lambda_\pi^*\over \Lambda_\pi}\cong {m_\rho^*\over m_\rho}\, .
\ee
Whereas the kinetic energies $T_{\rho\pi}$ in eq.~(2.12)
would seem to increase $\Lambda_\pi$, in fact in our picture
there are two pions, the $\pi$ and $\pi'$. The latter, when treated
as a correlated ($\rho,\pi$) system, coupled to
the quantum numbers of the $\pi$, has a broad mass distribution
starting at $m_\rho+m_\pi$. The elementary $\pi$
and the $\pi'$ mix pushing the $\pi$ down and the $\pi'$ up, in energy.
It is clear that the main player in the pion vertex function
is the mass of the $\rho$--meson.

We do not yet have a detailed description of the $\pi'$-meson, but
much of its mass must come from that of the $\rho$-meson.
We take $m_{\pi'}$ and $\Lambda_{\pi'}$ to scale as in eq.~(1.1),
although the scaling of the $\Lambda_{\pi'}$ has little effect.
The scaling of $f_{\pi'NN}$ is assumed to be the same as for $f_{\rho NN}$.

Finally we have to scale the mass of the $\Delta(1232)$ isobar.
It is known, e.g.
from inclusive electron scattering experiments, that the mass difference
between the nucleon and the $\Delta$ does not significantly change in the
nuclear medium. Therefore we keep this difference constant in our
calculations:
\be
m_{\Delta}^* - m_N^* = m_{\Delta} - m_N .
\ee

 From the above discussion, once the density dependence of $m^*_N$ is
known, the medium modification of all the other quantities can be derived.
For $m^*_N$ we shall assume a linear density dependence as
\be
{m_N^*(\rho) \over m_N}=1-0.3{\rho\over\rho_0}
\ee
which is adjusted to a value of $0.7$ at saturation density as shown in
Fig.~3 (for an extensive discussion of the nucleon effective mass in nuclei
see \cite{mcah}).  Brown and Rho \cite{brgm1} find, from the measured
isovector exchange current in $^{209}$Bi, a value of 0.75 for $m_N^*/m_N$
averaged over the nucleus, which is consistent with (2.15).

Whereas $g_A=1.26$, the in-medium coupling is roughly
\be
g_A^* \simeq 1
\ee
as is inferred from the quenching of magnetic moments and
$M1$ transitions, as well as the missing Gamow-Teller strength
\cite{gcao,ostf}. The renormalization of $g_A$ has two sources:
a screening through virtual $\Delta$-hole states \cite{rowo} and
second-order mixing of nucleonic excitations, chiefly through the tensor
force \cite{attk}. In our model the density dependence of $g^*_A$ is
determined, however, from the assumption that the ratio
$f_{\pi NN}/m_\pi$ remains fixed. Then using eq.~(2.6) as well as
the Goldberger-Treiman relation (2.11) gives that
\be
g_A^*=g_A\biggl({m^*_N\over m_N}\biggr)^{2/3}
\ee
(Fig.~3). At saturation density this yields $g_A^*/g_A=0.99$ in close
agreement with (2.16). From eq.~(2.7) the density dependence of
the $\rho$-meson mass is also determined (Fig.~3) and
we obtain a value of $0.79$ for $m_\rho^*(\rho_0)/m_\rho$. This is close to
QCD sum rule calculations \cite{bger,hlts}.\footnote{Hatsuda and Lee
\cite{hlts} give 0.82 as their central value. Chanfray and Ericson
\cite{cegm} have shown  that exchange current type processes involving the
virtual pion field decrease
the quark condensate by a factor $10-20\%$ over that of Hatsuda and Lee.
Birse and McGovern \cite{bmmj} show that with proper calculation of
the symmetry breaking matrix element, those result in an enhancement of chiral
symmetry breaking in nuclei.}

\singlespace
\section{The Quasifree Polarization-Transfer in the
(p,n) Reaction at 495 MeV}

\setcounter{equation}{0}

\doublespace
The $(\vec{p},\vec{n})$ polarization experiments aim at extracting the
ratio of the spin-longitudinal to spin-transverse response function
\be
{R_\parallel(q,\omega)\over R_\perp(q,\omega)}=
{\sum_{n\neq 0}|\bra{0}{\cal O}_\parallel\ket{n}|^2\delta(\omega-E_n)
\over
\sum_{n\neq 0}|\bra{0}{\cal O}_\perp\ket{n}|^2\delta(\omega-E_n)}
\ee
where
\be
{\cal O}_\parallel=\sum_{i=1}^A\bsigma_i\!
\cdot\!{\bf q}\tau_i^-e^{i{\bf q}\cdot{\bf r}_i}\, ;\qquad
{\cal O}_\perp=\sum_{i=1}^A\bsigma_i\!
\times\!{\bf q}\tau_i^-e^{i{\bf q}\cdot{\bf r}_i}
\ee
from a combination of spin-transfer coefficients. Measurements have been
performed  in $^{12}$C and $^{40}$Ca at a fixed
angle of $18^\circ$, corresponding to a peak
three-momentum transfer of 1.72 fm$^{-1}$, so as to maximize the
difference between $R_\parallel$ and $R_\perp$ expected from  the
 standard treatment of the response functions. The ratio,
however, is found to be essentially unity for projectile energy
losses below the quasielastic peak \cite{mjbc}. In our picture of
dropping masses this can be largely explained. We will sketch below the
recent work of Brown and Wambach \cite{bwgj}.

When looking at differences in the two-body interaction that cause a
deviation of $R_\parallel/R_\perp$ from unity, it is clear that
these can only come from the tensor part, since
\ba
V_\parallel=V_{\rm central}+2V_{\rm tensor}\nonumber\\
V_\perp=V_{\rm central}-V_{\rm tensor}   .
\ea
For simplicity, let us consider the $\pi$+$\rho$ exchange neglecting
form factors (the full $HT$ interaction leads to very similar conclusions).
In the static limit
\be
V_{\rm tensor}(q) =\Biggr\{ -{f_{\pi NN}^2 \over m_\pi^2}{q^2
\over(q^2+m_\pi^2)}
+{f_{\rho NN}^2\over m_\rho^2}{q^2\over(q^2+m_\rho^2)}\Biggr\}
S_{12}(\hat{\bf q})\mbox{\boldmath$\tau$}_1\cdot \mbox{\boldmath $\tau$}_2.
\ee
where $S_{12}(\hat{\bf q})=\mbox{\boldmath$\sigma$}_1\cdot\hat{\bf q}
\mbox{\boldmath $\sigma$}_2\cdot\hat{\bf q}
- 1/3(\mbox{\boldmath$\sigma$}_1\cdot\mbox{\boldmath$\sigma$}_2)$. A way
of  interpreting the experiment is to say that the tensor force is
essentially zero at the momentum transfer $q=1.72 fm^{-1}$.
We adopt as ratio of coupling constants \cite{bbns}
\be
{f_{\rho NN}^2\over m_\rho^2 }= 2{f_{\pi NN}^2 \over m_\pi^2 }
\ee
which, within a few percent, is that given by the HT interaction at the
relevant momentum transfer.
Since $m_\pi^2 \ll q^2 \ll m_\rho^2$, the requirement $V_{\rm tensor}=0$
implies that
\be
{f_{\rho NN}^2/m_\rho^2 \over f_{\pi NN}^2/m_\pi^2}
\left( {q^2 \over m_\rho^2} \right)\cong 1.
\ee
Using $m_\pi$ and $m_\rho$, however, from the Particle Data book, this
ratio is $0.4$ instead of unity.

\noindent
{}From eqs.~(2.9) and (3.6) we find that the condition for zero net tensor
force at $q=1.72fm^{-1}$ becomes\footnote{The loop correction $g_A$ was
not introduced in ref.~\cite{bwgj}.}
\be
\left( {m_\rho \over m_\rho^* } \right)^4 {g_A \over g_A^*} =2.5,
\ee
or, assuming $g_A^*=1$,
\be
\left( {m_\rho \over m_\rho^* } \right)^4 =2.
\ee
This gives
\be
m_\rho^* /m_\rho =0.84
\ee
which is slightly larger than the value of 0.79 at saturation density, found
above.

One should note that the $(p,n)$ reaction at 500 MeV is strongly
surface dominated and therefore the nucleus is probed at a lower density.
In the calculations of ref.~\cite{bwgj} this was taken into account in a
semiclassical description, including distortion
effects, which amounts to an averaging of the {\it local} nuclear matter
response functions with a weight factor $F$ as
\be
R^A_{\parallel,\perp}(q,\omega)=\int d^3r R^{\rm NM}_{\parallel,\perp}
(q,\omega;\rho(r))F(r).
\ee
The weight factor can be evaluated using the Eikonal approximation
\cite{cgha}. The average density $\ave{\rho}=\int d^3r\rho(r)F(r)/
\int d^3rF(r)$ turns out to be $0.35 \rho_0$ at which $m^*_\rho/m_\rho=0.93$.
It is therefore expected that the tensor interaction is reduced but does
not completely vanish. As shown in Fig.~4 this is born out of the more
complete calculation using the full {\it HT} interaction as well as all
the medium modifications discussed in sect.~2.
The results  (full lines) for $^{40}Ca$ at $q=1.72$ fm$^{-1}$ (left panel)
and a more recent measurement for $^{12}C$ at
$q=1.2$ fm$^{-1}$ \cite{mjbc1} (right panel) give an improvement for energies
below the quasielastic peak as compared to a standard RPA treatment
(dashed-dotted lines). Our model should be most reliable, however, near the
quasielastic peaks of $\omega\simeq 82 MeV$ for $q=1.72 fm^{-1}$
and $\omega\simeq 55 MeV$ for $q=1.2 fm^{-1}$ \cite{mjbc1},
including a $Q$-value of $\sim 20$ MeV for the $(p,n)$ reaction.
For substantially lower $\omega$ the finite-nucleus
collective excitations have to be treated explicitly. For excitation
energies above the
quasielastic peak, on the other hand,  one should bear in mind that
contributions from two-step processes become significant \cite{bsgo}.
Double scattering contributions to the spin observables are not well
understood at present.
For $q=2.5 fm^{-1}$ Brown and Wambach \cite{bwgj} predicted a ratio below
unity, which has been experimentally confirmed very recently by Taddeucci
{\it et al.} \cite{mjbc1}.
However, at this large momentum transfer our results are
very sensitive to the exact scaling behavior of the $\pi'$ which is
presently worked out within the microscopic model of ref.
\cite{jdhp}. We therefore postpone the discussion of the $q=2.5 fm^{-1}$
data until these investigations are finished.

The authors of ref. \cite{mjbc1} come to the conclusion that the ratio
$R_\parallel / R_\perp$ remains small because of an unexpected enhancement
of the transverse response rather than a non-enhancement of the longitudinal
one. There is no theoretical explanation for such an effect. However, the
data analysis strongly depends on the treatment of the distortion which
enters in terms of an effective number of neutrons. Because of the short
range character of the interaction in the transverse channel it is quite
reasonable that the distortion is much larger in this channel than in the
longitudinal one. Of course, this would also change the ratio.

Our model contains as an essential ingredient the change of
the nucleon effective mass $m_N^*$ with density. It is useful
to single out this effect in order to make contact with the recent
relativistic calculations of Horowitz and Piekarewicz \cite{hpcj},
since the nucleon effective mass enters much in the same way as in the
relativistic treatment. A drop of $m_N^*$ has two effects: (1) the density
of particle-hole states is decreased. This shifts the position of
the quasielastic peak to higher energies and broadens it. At the same
time the longitudinal and transverse correlations are weakened because of
an increase of the particle-hole energies. (2) more importantly $g'_{NN}$
is increased. Recall that $g'_{NN}$ receives a significant
contribution from $\rho$ exchange.
At zero density and momentum transfer we find the contribution
to be $g_{NN}^{\prime (\rho)} = 0.28$
while $\pi$ and $\pi'$ contribute with 0.08 and 0.21 respectively.
By using eq.~(2.8) one can relate the increase of the $\rho$ contribution
to $m^*_N$:
\be
g_{NN}^{\prime(\rho)}(\rho) \cong g_{NN}^{\prime (\rho)}(\rho=0)
\left( {m_N \over m_N^*}
\right)^2   .
\ee
Using $m^*_N/m_N=0.7$ appropriate for $\rho_0$ we find
$g_{NN}^{\prime (\rho)}(\rho_0)=0.57$.
This would increase $g'_{NN}$ to a value of 0.86 at nuclear matter density
which is very similar to the $g'_{NN} = 0.9$ of Horowitz and Piekarewicz
although they assign only a value of 0.3 from $\rho$-exchange. However, in
our model the $\pi'$ scales in the same way as the $\rho$ and this brings
$g'_{NN}$ up to 1.08 (see Fig. 5). On the other hand this additional short
range correlation due to the $\pi'$ is canceled or even overcompensated by
an increase of the $\pi'$ contribution to the tensor force. Therefore,
compared with the results of Horowitz and and Piekarewicz we find a
somewhat weaker effect of the scaling on the longitudinal-transverse ratio.
Preliminary results by Janssen \cite{jdhp} seem to indicate that the
$\pi'NN$ coupling constant coming out of a more microscopic calculation
will be much weaker than the HT value. This would considerably improve
our results.

So far only the drop of the nucleon mass entered into our argumentation.
As can be seen by comparing the solid and dashed lines in Fig.~4
dropping $m_\rho$ and the other properties as described in section 2
in addition to $m_N$ does not have a large effect. One
concludes that the quasielastic polarization transfer experiments are
most sensitive to the change in the nucleon effective mass. In the
deep-inelastic scattering experiments the situation is quite different.

\section{The (Lack of) EMC and Drell-Yan Effects}

\setcounter{equation}{0}

We shall discuss here the region of $0.1<x<0.3$ where pion
enhancement effects were supposed to be and where they weren't \cite{bfsg}.
Below $x=0.1$ there is shadowing, an interesting phenomenon in
its own right. In any description which fits the shadowing,
(see, {\it e.g.} the 'reggeized' discussion of Brodsky and Lu \cite{blsh})
and preserves the momentum sum rule, there will be some small overshoot
above $x=0.1$. It is argued that this overshoot concerns valence quarks
\cite{fsll}, $i.e.$ it enters only into the EMC effect. We also will
not discuss  the dip in the EMC effect in the region of $x\sim 0.5-0.6$.
There are parameterizations of this dip \cite{crrf}-\cite{cfel} in terms
of rescaling.

In the region $0.1<x<0.3$ the EMC as well as the Drell-Yan ratio are
sensitive to the sea quark distributions in the nucleus.
We consider a two phase model where the nucleon is made up of a bare quark
core and a second component consisting of virtual meson-baryon pairs.
Therefore the structure function $F_2$ reads:
\be
F_2^N(x) = Z_N \{ F_2^{core}(x) + \sum_i (\delta F_2^{B_i/N}(x)
                                        + \delta F_2^{M_i/N}(x))\} .
\ee
Here $Z_N$ denotes a wave function renormalization constant, on which we
comment below. The sum in the bracket may run over all meson-baryon
decompositions of the nucleon. In deep-inelastic scattering processes the
virtual photon couples to the core as well as to the recoil baryon
(described by
$\delta F_2^{B_i/N}$) and the meson (described by $\delta F_2^{M_i/N}$).
The most important example for the latter case is the Sullivan process
\cite{sjdu} where the photon couples to the pion cloud (Fig. 6(a)).
The corresponding contribution to the structure function of the nucleon is:
\be
Z_N \delta F_2^{\pi / N}(x) =
Z_N \int_x^1 dy f^{\pi/N}(y) F^\pi_2(\frac {x}{y})
\ee
with
\be
f^{\pi/N}(y) = \frac{3}{16\pi^2} \; g^{2}_{\pi NN} \; y
\int ^{\infty}_{m_{N}^{2} y^{2}/(1-y)} dt \; t \frac{|\Gamma_{\pi NN}(t)|^{2}}
{(t +
m_{\pi}^{2})^{2}}
\ee
being the probability of finding a pion in the nucleon which carries the
plus-momentum fraction
\be
y = \frac{p_\pi^o + p_\pi^3}{m_N} .
\ee
The renormalization constant $Z_N$ (which is missing in the original paper
by Sullivan \cite{sjdu}) normalizes the total probability of finding the
nucleon in one of the two phases to unity. The importance of this constant
in connection with number sum rules has been shown
by Szczurek and Speth \cite{ss93}. It is given by
\be
Z_N = (1 + \sum_i \int_0^1 dy f^{M_i/N}(y))^{-1} ,
\ee
where $f^{M_i/N}$ is the distribution function for the meson $M_i$,
analogous to eq. (4.3). Taking into
account a large set of processes the authors of ref. \cite{ss93}
find $Z_N \simeq 0.6$ for a cut-off which roughly corresponds to
a monopole form factor with $\Lambda = 800 MeV$. We adopt this value
for our calculations.

In the nuclear medium analogous relations hold. We assume that the
structure function of the mesons and the baryon cores remain unchanged
whereas the meson distribution functions $f^{M_i}$ and the corresponding
ones for the recoil baryon have to be modified. As
shown in ref. \cite{MB} the modification of the baryon part can be absorbed
in the Fermi motion of the nucleons. Furthermore, in the kinematical region
we are interested in ($x \leq 0.3$), the only relevant contribution comes from
the pion and we can neglect the change of the other meson distribution
functions. Thus the structure function $F_2$ of a nucleon in the nuclear
medium becomes:\footnote{For simplicity we discuss only isospin averaged
structure functions. In nuclei with neutron excess, like $^{56}Fe$, the
pion cloud contains more $\pi^-$ than $\pi^+$ mesons, i.e. a small amount
of negative charge is transferred from the nucleons to the pion cloud. This
effect is properly taken into account in our numerical calculations although
it is almost negligible. For details see ref. \cite{MB}.}
\be
F_2^{N/A}(x) = \int_x^A dz f^{N/A}(z) F_2^N(\frac {x}{z}) +
\int_x^A dy (Z_A f^{\pi/A}(y) - Z_N  f^{\pi/N}(y)) \; F^\pi_2(\frac {x}{y}) .
\ee
The function $f^{N/A}(z)$ in the first integral describes the nucleon
distribution due to Fermi motion. Since Fermi motion is not very important
at small values of $x$ the main effect comes from the change of the pion
distribution function which gives rise to the second integral. Up to this
point everything is like in the conventional pion excess model. How\-ever,
in eq.~(4.6) the pion distribution functions $f^{\pi/N}$ and $f^{\pi/A}$
are multiplied by the normalization factors $Z_N$ and $Z_A$, respectively.
Assuming that we can neglect the change in the distribution functions for
mesons other than the pion it follows from eq. (4.5) and the analogous
equation for $Z_A$:
\be
Z_A^{-1} = Z_N^{-1} + \int_0^1 dy (f^{\pi/A}(y) - f^{\pi/N}(y)) ,
\ee
{\it i.e.} the normalization factor $Z_A$ decreases with an increasing
distribution function $f^{\pi/A}$. Thus the second integral in eq. (4.6)
is much smaller (about a factor of $Z_N^2$) than it would be without
normalization factors.

In our model the in-medium distribution function is given by
\be
f^{\pi/A}(y) = \frac{3}{16\pi^{2}} \; g^{2}_{\pi NN} \; y
\int ^{\infty}_{m_{N}^{2} y^{2}} dt \int^{\frac {t-m_N^2 y^2}{2m_Ny}}_0
d\omega \;t \frac{|\Gamma_{\pi NN}(t)|^2
R_\parallel(\omega ,\sqrt{t})}{(t + m_{\pi}^{2})^{2}} ,
\ee
with $R_\parallel(\omega ,q)$ being the non-relativistic
spin-longitudinal response function for nuclear matter. It describes
Pauli blocking as well as
rescattering corrections (see Fig.~6(b)). Pauli
blocking leads to a depletion in the mean number of pions per nucleon
as compared to the free nucleon which is overcompensated by
the rescattering, chiefly through $\Delta$-hole excitations, giving
a net pion excess.

As in eq.~(4.3), $t$ is the (space-like) four-momentum transfer
$t = {\vec q}^{\,2} - \omega^2$.
The integration limits for $t$ and $\omega$ follow directly from
eq. (4.4)
(with $p_\pi^o = - \omega$ and $|\vec p_\pi| = |\vec q|$).
This is different from the distribution functions which can be found in the
literature \cite{etma} \cite{bbmr}, where the three-momentum transfer is the
integration variable (leading to $\omega_{max} = |\vec q| - m_N y$) as well
as the second argument of $R_\parallel$. In the non-relativistic regime,
{\it i.e.}
when the main contributions to the integral come from regions with
$\omega \ll |\vec q|$, both prescriptions become identical. In addition,
however, eq.~(4.8) has the correct relativistic low density limit:
\be
\lim_{k_F \rightarrow 0} f^{\pi/A}(y) = f^{\pi/N}(y) ,
\ee
which is not the case for the function given in refs. \cite{etma}
and \cite{bbmr}. Since both functions, $f^{\pi/A}$ and $f^{\pi/N}$, enter
into eq. (4.6) we prefer the more consistent prescription eq.~(4.8).
In our final results this enhances the pion contribution by a few percent.

For the Fe nucleus we choose an average density of $\ave{\rho}=0.87 \rho_0$
corresponding to $k_F=260$ MeV \cite{mejo}\footnote{Of course, the Drell-Yan
(and EMC) experiments see a higher $\ave{\rho}$ than the polarization
transfer, because in the latter case the projectile is affected by the
strong interactions.}.
With no medium modified masses we obtain for the first moment of the pion
distribution function $M_1^{\pi/A} = \int dy f^{\pi/A}(y)$ a value of 0.70
at this density. This has to be compared with the free value of $M_1^{\pi/N}
= 0.41$ (see also Fig.~7).
However, because of the normalization factors this enhancement
has much less influence on the EMC and Drell-Yan ratios than expected in the
past. The corresponding predictions are displayed as the dashed lines in
Figs.~8 and 9. In the region $0.1 \le x \le 0.3$ there is no significant
deviation of the predicted EMC ratio from the data. Of course, this does not
mean that the pion field {\it is} amplified. Rather, since the more sensitive
Drell-Yan data remain strongly overestimated, we are still led to the
conclusion that the standard picture fails.
This cannot be reconciled by changes in the key parameter
$g'_{N\Delta}$ which would have to be chosen unrealistically large. It also
seems implausible that more sophisticated many-body approaches will cure
this problem.

With dropping masses, coupling constants and formfactors several
modifications occur.
Brown, Li and Liu \cite{bllg} pointed out that the nucleon
effective mass
$m^*_N$ rather than $m_N$ should be used at the soft $\pi NN$ vertices
in Fig.~6. The nucleon mass enters at two places into the derivation
of the Sullivan formula (eq.~(4.3)).
The first place is the spin-isospin current which mixes the large and
the small spinor components of the nucleon. Secondly the energy of the
virtual pion, as a function of its momentum, is determined from the
on-shell condition for the initial and the final nucleon. In both cases
the nucleon effective mass should be used.

The modification of the Sullivan formula
due to Brown/Rho scaling can be obtained most easily by replacing  all
properties on the r.h.s. of eq.~(4.2)  and in eq.~(4.3)
by the scaled ones. This also includes the variables $x$ and $y$ which
have to be replaced by $x^* = \frac{Q^2}{2m_N^*\nu}$ and
$y^* = \frac{p_\pi^o + p_\pi^3}{m_N^*}$. Substituting back to the
variable $y = \frac{m_N^*}{m_N} y^*$ we find
\be
\delta {F_2^{\pi / N}}^*(x) = \int_x^{\frac{m_N}{m_N^*}} dy
{f^N_\pi}^*(y) F^\pi_2(\frac {x}{y})
\ee
with
\be
{f^N_{\pi}}^*(y) = \frac{3}{16\pi^2} \; {g^*}^2_{\pi NN}\;
(\frac{m_N}{m_N^*})^2 \; y
\int ^{\infty}_{m_{N}^{2} y^{2}/(1-\frac{m_N}{m_N^*}y)}
dt \; t \frac{|\Gamma^*_{\pi NN}(t)|^{2}} {(t + m_{\pi}^{2})^{2}}  .
\ee
Note that the coupling constant $g^*_{\pi NN}$ comes together with a
factor $\frac{m_N}{m_N^*}$. As we have argued in sect. 2 this combination
is density independent as long as we keep $f_{\pi NN}/m_\pi$ constant.
Compared with eq. (4.3), eq. (4.11)
leads to a reduced number of pions: The first reason is the smaller
cutoff $\Lambda_{\pi NN}^*$ in the $\pi NN$ form factor. The second
reason is the enhanced lower limit of the t-integration.

Including nucleon-hole and $\Delta$-hole rescattering diagrams (Fig. 6(b))
amplifies the pion field again. Because of the stronger
short-range repulsion the effect is
smaller than without scaling but it is still present. The general
behavior can be seen from Fig. 7 where the first moment of the pion
distribution
function, $M_1^{\pi/A}$, is plotted as a function of density.
Whereas $M_1^{\pi/A}$ is strongly enhanced in the standard RPA calculation
(dashed-dotted line) the scaled result (solid line) comes quite close
to the free nucleon value (dotted line). At $\rho = .87 \rho_0$
which corresponds to the averaged density of the $Fe$ nucleus we find
$M_1^{\pi/A} = 0.70$ without and
$M_1^{\pi/A} = 0.39$ with Brown/Rho scaling which
has to be compared with $M_1^{\pi/N} = 0.41$ for the free nucleon . The dashed
line shows the result of the scaled first-order calculation  (eq. (4.9)).

As can be expected from these results
we almost produce a null effect in the Drell-Yan
as well as in the EMC experiments (full lines of Fig.~8 and Fig.~9). This can
be seen by comparison with the dotted lines, which show the result of a
calculation with the pion contribution (second integral of eq.~(4.6)) being
switched off.
The EMC data in the $x$-region of interest are even
somewhat underestimated. To obtain the change in the quark distributions and
the nucleon structure function (eq.~(4.6)) we have employed the quark
distributions of the free nucleon and the pion by Owens \cite{dodj}.
Other parameterizations
\cite{bjad}-\cite{bbet} yield basically the same result. In the EMC
calculation nuclear separation energy effects and Fermi motion have been
put in as in refs. \cite{bllg} and \cite{llbg}. As discussed
by Li, Liu and Brown \cite{llbg}, only part of the dip at larger $x\sim 0.6$
is explained by binding energy effects, once the proper normalization
for the baryon number is used.
As mentioned above, the enhancement in the region of
$x=0.1$ to $0.2$ seems to be in the valence quarks and such an
enhancement can sensibly come from antishadowing or any description of the
shadowing which preserves the momentum sum rule.

The pion enhancement in both Drell-Yan and EMC experiments involves
convolutions over a fairly wide region of momenta. In this sense, they
are less specific than the Los Alamos polarization transfer experiments.
While in the former the dropping nucleon mass and the resulting density
dependence of $g'_{NN}$ is the key physical effect, in agreement with the
findings by of Horowitz and Piekarewicz \cite{hpcj}, the deep inelastic
experiments are much more sensitive a change of the $\rho$-meson mass,
chiefly through the softening of the $\pi NN$ vertex in the medium.
The Drell-Yan data cannot be described if this softening
is not taken into account.

\section{Summary}

\setcounter{equation}{0}

We have analyzed three recent experiments which have looked at effects
of an enhancement of the virtual pion field in nuclei. The negative
outcome can be understood from a perspective of partial restoration
of chiral symmetry with density which reflects itself in a drop of
hadron masses, especially the nucleon and $\rho$-meson mass. Following
recent developments in the nucleon-nucleon interaction which try to
reconcile the sea quark distribution in the nucleon with the low-energy
properties of the two-nucleon system, as well as employing microscopic
calculations of the $\pi NN$ form factor we are able to explain the apparent
lack of pion enhancement in nuclei. Thus the large discrepancies
\cite{bfsg} between  the conventional theory and experiment are removed.

\begin{center}
\bf{Acknowledgement}
\end{center}
We would like to thank Leonid Frankfurt, Gerry Garvey and Mark Strikman
for many stimulating discussions. This work was supported in part by
DOE grant DE-FG 0188ER40388 and NSF grant PHY-89-21025. One of us (M.B.)
was supported in part by the Feodor Lynen program of the Alexander von
Humboldt foundation.

\newpage
\pagestyle{empty}
\begin{center}
\centering{\bf{\large Figure Captions}}
\end{center}
\vspace{1.3cm}
\begin{itemize}
\item[{\bf Fig.~1}]
The momentum dependence of the Fermi liquid parameters
$g'_{NN}$, $g'_{N\Delta}$ and $g'_{\Delta\Delta}$ at zero density
deduced from the
potential \cite{htka} by using a realistic two-body correlation function.
For $g'_{N\Delta}(0)$ and $g'_{\Delta\Delta}(0)$ the classical Lorentz-Lorenz
value was taken in agreement with the data analysis by Thies \cite{tmhi}
and Johnson \cite{jmoh}.

\item [{\bf Fig.~2}]
The vertex function for the $\pi NN$ interaction. Here we have displayed only
the simplest contribution. Because of self interactions, a large number of
higher-order diagrams are possible \cite{jdhp}.

\item [{\bf Fig.~3}]
The density dependence of the key quantities in our description. Assuming
a linear dependence for $m^*_N(\rho)/m_N$ such that $m^*_N(\rho_0)/m_N=
0.7$ the effective $\rho$-meson mass is fixed by eq.~(2.7) while
$g^*_A(\rho)$ is determined by the requirement that the $\pi NN$ coupling
$f_{\pi NN}/m_\pi$ in the medium is the same as in free space. The
empirical value of $m^*_N$ is taken from ref.~\cite{mcah} while that
of $g_A^*$ is estimated from \cite{ostf}.

\item [{\bf Fig.~4}]
The ratio of spin-longitudinal to -transverse response functions in
$^{40}$Ca as a function of excitation energy and at fixed momentum transfer
$q=1.7$ fm$^{-1}$ (left panel). The dashed line gives the result of a
standard RPA treatment with only nucleon effective mass, while the full
line includes the effects of medium-dependent nucleon mass and meson masses.
The dashed-dotted line displays the result
without any effective mass. The data (measured at a fixed angle of $18^\circ$
corresponding to $q=1.7$ fm$^{-1}$ at and below
the quasielastic peak $\omega = 80 MeV$)
were taken from refs.~\cite{mjbc} (open circles) and \cite{mjbc1} (solid
circles).
The right panel displays the predictions for a momentum transfer of $^{12}C$
at 1.2 fm$^{-1}$ recently measured at LAMPF \cite{mjbc1}
($\theta = 12.5^\circ$).
The labeling of the curves is the same as in the left panel.

\item [{\bf Fig.~5}]
left panel: the density dependence of the Fermi liquid parameters $g'(0)$
after inclusion of medium-modified masses. The empirical values for $g'_{NN}$
and $g'_{\Delta\Delta}$ (see text) are also given, including their
uncertainties.\\
right panel: the momentum dependence of the Fermi liquid parameters $g'$
at saturation density, $\rho_0$.

\item [{\bf Fig.~6}] (a) Deep inelastic scattering off a pion
in lowest order.
(b) Rescattering correction to (a).
The shaded areas in the bubbles are vertex corrections which introduce
the local field correction $g'$; nucleon-hole and $\Delta$-hole
intermediate states are included similarly
as in the works of refs. \cite{etma} and
\cite{bbmr}.

\item [{\bf Fig.~7}]  The first moment
$M_1^{\pi/A} = \int_0^1 dy f^{\pi/A}(y)$
of the pion distribution function as a function of density. The dotted
line indicates the free nucleon value $M_1^{\pi/A} = .41$. The standard RPA
result without Brown/Rho scaling is represented by the dashed-dotted line
while the solid line corresponds to the calculation with scaling. The
dashed line shows the result of the scaled first-order calculation
(no rescattering), corresponding to eq. (4.11).

\item [{\bf Fig.~8}] The EMC ratio: The dashed
line gives the result of a conventional RPA treatment without medium
modifications of the hadron masses, while the full line includes those
effects. Wave function renormalization constants have been used in both
cases as described in the text. Switching off the pion contribution
(second integral of eq.~(4.6)) one obtains the result represented by the
dotted line.
The calculations have been performed at
$\rho = 0.87\rho_o$ which corresponds to the average density of $^{56}Fe$.
The data were taken from refs.~\cite{apma} ($^{40}Ca$) and \cite{argr}
($^{56}Fe$).

\item [{\bf Fig.~9}] The Drell-Yan ratio: The dashed
line gives the result of a conventional RPA treatment without medium
modifications of the hadron masses, while the full line includes those
effects. The dotted line represents the result of a calculation where
the pion contribution has been switched off.
The data were taken from ref.~\cite{adml} ($^{56}Fe$).

\end{itemize}

The figures can be ordered from the authors via
Michael.Buballa@sunysb.edu $\;\;.$

\end{document}